# Singlet Fission among Two Single Molecules

Sumanta Paul[1,†], Oleksandr Yampolskyy[1,†], Zehua Wu[1], Klaus Müllen[1,2], Thomas Basché[1,*]

[1]Department of Chemistry, Johannes Gutenberg-Universität, Mainz, Germany

[2]Max Planck Institute for Polymer Research, Mainz, Germany

*Corresponding author. Email: thomas.basche@uni-mainz.de

[†]These authors contributed equally.

## Abstract

Singlet fission (SF) is a photophysical process where a singlet excitation generates two triplet excited states, enhancing exciton multiplication potentially useful for solar energy conversion. Since SF typically outcompetes radiative decay, single molecule studies of SF have remained elusive. Here, we present single molecule spectroscopy of a terrylenediimide (TDI) dimer at room and cryogenic temperatures. By analysing the stream of photons emitted by single dimers, the rates of formation and decay of SF-born triplet states were determined. We report considerable static and dynamic heterogeneities of the SF process which are reflected in broad rate distributions as well as the occasional occurrence of delayed fluorescence and rate fluctuations during spin evolution. Cryogenic experiments point to the formation of a coherent multiexciton superposition state which decays into the singlet exciton from which a correlated triplet pair evolves. Our results establish single molecule spectroscopy as a new avenue into mechanistic details of the SF process which often are drowned by ensemble averaging.

# Introduction

In singlet fission (SF) the absorption of a single photon in molecular assemblies results in the conversion of a singlet exciton into two lower-energy triplet excitons, each localized on adjacent chromophores[1,2]. SF has garnered significant attention for its potential to enhance the theoretical Shockley–Queisser efficiency limit of single-junction solar cells from 33% to 45%, as well as its emerging applications in quantum information science [2–6]. SF requires specific conditions: the energy of the lowest singlet excited state (E ($S_1$)) should be roughly twice that of the lowest triplet excited state (E($S_1$) ≥ 2E ($T_1$)), and the energy of the second triplet state (E($T_2$)) should exceed 2 E($T_1$)[1,2,7–9]. Spin conservation entails that the initial product of SF is a correlated triplet pair with singlet spin multiplicity, $^1$(TT)[1,2,8,9]. The formation of $^1(T_1T_1)$ can proceed through either a coherent or an incoherent SF mechanism. In the coherent pathway, a superposition of the delocalized singlet state $^1(S_1S_0)$ and $^1(T_1T_1)$ forms immediately after excitation and subsequently dephases to yield $^1(T_1T_1)$[1,2,10–15]. In the incoherent mechanism, $^1(T_1T_1)$ may form directly from $^1(S_1S_0)$ or via a charge-transfer (CT) state that mediates the transition[1,2,11,16,17]. Following this first step, the interconversion of $^1$(TT) with other triplet-pair states of different spin multiplicities seems to be characteristic for many SF systems. Hereby, a complex ensemble of multi-excitonic triplet-pair states may be involved, distinguished by their degree of exchange coupling, which depends on the orbital overlap and interchromophore distance[18–24].

Considering molecular dimers, the SF process has been modelled by the following simplified scheme[9,11,19–23,25,26]:

$$[\,^1(S_1S_0) \rightleftharpoons \,^1(T_1T_1)] \rightleftharpoons \,^m(T_1 \cdots T_1) \rightleftharpoons (T_1 + T_1) \qquad (1)$$

While the $^1(T_1T_1)$ state may emerge at an ultrafast timescale from $^1(S_1S_0)$ there may be an equilibrium between the two states, or a coherent superposition state is formed as indicated by the square brackets in Equation 1. Since spatial dissociation of the triplets is precluded, spin evolution can lead to a spin-entangled mixed state $^m(T_1 \cdots T_1)$ of singlet and quintet character as has been reported for several molecular species[19–22,27]. Eventually, $^m(T_1 \cdots T_1)$ may decay into two non-interacting triplet states ($T_1 + T_1$).

Recently, rylene diimides – in particular covalently linked dimers of terrylene diimide (TDI) - have emerged as promising model systems for investigating the SF mechanism[28–34]. Such dimers enable modulation of electronic coupling as well as exchange coupling through rational linker design. Notably, SF in TDI dimers with covalently linked and slip-stacked geometries was found to initially involve a mixed coherent superposition state with contributions from the $^1(S_1S_0)$, $^1(T_1T_1)$ and a charge transfer state[32–34]. Similar ultrafast state mixing was also reported for a collinear TDI dimer, where two TDI molecules are directly connected at their imide nitrogens (Fig. 1b)[30,31]. In this system contributions from a CT state could be neglected and the coherent superposition state with mixed $^1(S_1S_0)$ and $^1(T_1T_1)$ character evolved within several tens of femtoseconds. The coherent superposition state is driven by nonadiabatic vibronic coupling involving low-frequency $^1(S_1S_0)$ modes and high-frequency $^1(T_1T_1)$ modes[31]. Subsequently, non-interacting triplets ($T_1 + T_1$) are formed via an intermediate $^m(T_1 \cdots T_1)$ state and then decay to the ground state[30].

Single-molecule spectroscopy studies of SF so far have remained elusive, most probably because SF typically is accompanied by low fluorescence quantum yields. Yet, such studies promise to deliver unique insights into static and dynamic heterogeneities of the SF process which often are drowned in the ensemble average. Moreover, the extremely low concentrations ($\approx 10^{-10}$ mol/ L) used in single-molecule studies largely prevent aggregation of extended π-systems. When looking at a single molecule undergoing transitions between optically bright and dark states, the stream of emitted photons contains information about the temporal succession of these transitions without the need of an external trigger. A powerful way to analyse the underlying kinetics is to measure the fluorescence intensity autocorrelation function which shows photon antibunching at short times and photon bunching due to bright-dark transitions[35–39].

Here, we employed bulk and single molecule spectroscopy to investigate intramolecular SF in the collinear TDI dimer (Fig. 1b). The favourable photophysical properties of TDI for single molecule spectroscopy at room and low temperature[40,41] and a fluorescence quantum yield in solution of roughly 50 % highly qualified this dimer for the intended study. The properties of the emissive state are in accordance with a delocalized $S_1$ exciton with a fluorescence lifetime containing contributions from superradiance[42,43] and SF. By analysing the fluorescence intensity autocorrelation function of single dimers, we find an overall SF rate which is almost 3 orders of magnitude larger than the intersystem crossing (ISC) rate in TDI monomers while the triplet decay rate basically remains unchanged. Together with the occasional observation of delayed fluorescence, the wide distributions of rate parameters reveal significant differences between individual dimers (static heterogeneity). Moreover, correlated fluctuations of the SF rate and the rate of delayed fluorescence point to slow spin evolution as a signature of dynamic heterogeneity. Strongly broadened excitation spectra of single dimers at 1.4 K point to the initial formation of a mixed $^1(S_1S_0)$ and $^1(T_1T_1)$ superposition state which decays on an ultrafast time scale. Assuming subsequent formation of the delocalized pure $^1(S_1S_0)$ state, rationalises slow overall SF as well as the large fluorescence quantum yield.

**Results**

**Bulk Absorption and Fluorescence Measurements.** The synthesis of the TDI monomer[44] and the TDI Dimer[30] followed procedures previously described in the literature (Supplementary Fig. 1-4). The steady-state absorption and emission spectra of the TDI monomer and dimer in toluene are presented in Fig. 1 (a) and 1 (b). The absorption spectrum of the TDI monomer features three well-resolved bands at 652 nm, 599 nm, and 552 nm, corresponding to the 0-0, 0-1, and 0-2 vibronic transitions, respectively. In case of the TDI dimer these bands are red shifted by about 12-13 nm. The fluorescence spectra of both species closely mirror their absorption spectra, displaying the same vibronic band signatures which are red shifted by 11-12 nm for the dimer (Table 1). Notably, the relative intensities of the 0-0 and 0-1 bands differ significantly between the dimer and monomer in both absorption and emission spectra (Fig. 1 (a) and (b)). In particular, the TDI dimer shows an increased ratio of the intensity of the 0-0 bands ($I_{0-0}$) to the 0-1 ($I_{0-1}$) bands compared to the monomer. The increase of the $I_{00}/I_{01}$ ratio in the dimer reports about exciton delocalization and signals the formation of a coherent state through electronic coupling.[42,43] In accordance with the SF literature, we have denoted this state with $^1(S_1S_0)$ keeping in mind that we are dealing with a delocalized $S_1$ exciton state. Because the HOMO and LUMO possess orbital nodes at the imide linkage, through bond conjugation can be largely ruled out. In addition, orbital overlap is negligible due to the almost orthogonal orientation of the TDI units. Assuming through space electrostatic dipole-dipole coupling (see Methods for details), the coupling strength V was estimated to be 240 cm$^{-1}$, which predicts a red-shifted dimer emission maximum at 674 nm, in good agreement with the experimental observation of 675 nm.

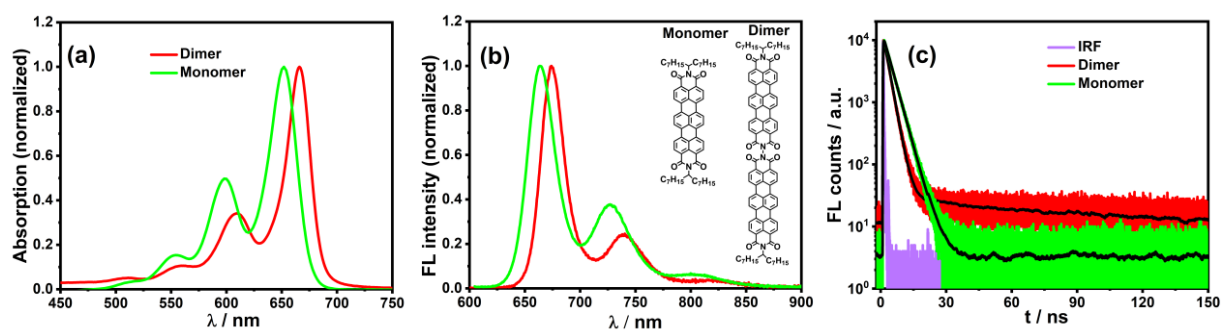

**Fig. 1 | Exciton coherence and delayed fluorescence in TDI dimers.** Normalized absorption (a) and fluorescence (b) spectra ($\lambda_{exc}$ = 594 nm) of the TDI monomer (c = $6.4 \cdot 10^{-7}$ mol/L) and dimer (c = $4.7 \cdot 10^{-7}$ mol/L) in toluene. The molecular structures of the monomer and dimer are shown in the inset of panel (b). Fluorescence decay curves (c) of the TDI monomer and dimer in toluene, recorded after excitation at 600 nm and monitored at their respective emission maxima. The black lines represent the decay models fitted to the data.

Fluorescence decay curves of the TDI monomer and dimer in toluene are shown in Fig. 1 (c). The TDI monomer exhibits a single exponential fluorescence decay with a lifetime of 3.4 ns (Table 1). In contrast, the TDI dimer displays a biexponential fluorescence decay, featuring a dominant short lifetime component ($\tau_{FL}$) of 2.2 ns (~95%) and a minor long lifetime component ($\tau_{del}$) of 80 ± 17 ns (~5%). The long lifetime component which was also found at the single molecule level (see below) is attributed to delayed fluorescence which was not reported in a previous investigation[30]. The substantial shortening of the short lifetime component from 3.4 ns in the monomer to 2.2 ns in the dimer is accompanied by only a slight decrease of the fluorescence quantum yield from 0.7 (monomer) to 0.64 (dimer). These numbers are likely influenced by the opposing effects of SF on the one side and electronic coupling on the other side. The non-radiative process of SF is expected to lead to a decrease of the fluorescence lifetime and quantum yield. While electronic coupling also leads to a decrease of the lifetime, this is due to an increase of the radiative rate (superradiance) which should be accompanied by an increase of the quantum yield.[43] The radiative rate is given by $k_{rad} = \Phi_{FL} \cdot k_{FL}$, with $k_{FL} = (\tau_{FL})^{-1}$. For the TDI monomer we obtain $k_{rad} = 2.1 \cdot 10^8$ s$^{-1}$. For the TDI dimer we find $k_{rad} = 2.9 \cdot 10^8$ s$^{-1}$, in good agreement with an independent estimate based on the coherence enhancement number $N_{coh}$ (see Methods for details).

The fluorescence decay rate is given by all processes which lead to the decay of the $S_1$ state. In case of the TDI monomer this reads: $k_{FL} = k_{rad} + k_{IC} + k_{ISC}$, with $k_{IC}$ and $k_{ISC}$ being the $S_1$-$S_0$ internal conversion rate and the intersystem crossing rate, respectively. Since $k_{ISC}$ is much smaller than all other rates (see below), it can be safely neglected here. For the TDI monomer we find $k_{IC} = 8 \cdot 10^7$ s$^{-1}$. Assuming that the IC rate does not change in the dimer, the radiative rate and IC rate add up to a value of $3.7 \cdot 10^8$ s$^{-1}$ which is smaller than the total fluorescence decay rate of the dimer, $k_{FL} = 4.5 \cdot 10^8$ s$^{-1}$. The difference ($8 \cdot 10^7$ s$^{-1}$) is assigned to the additional contribution from SF as will be corroborated and more rigorously quantified by the single molecule measurements.

**Table 1| Spectral and photophysical parameters of the TDI monomer and dimer.**

| System | Medium | $\lambda_{abs}^{max}$ nm | $\lambda_{FL}^{max}$ nm | $\tau_{FL}$ ns | $\tau_{del}$ ns | $\varphi_{fl}$ | $k_{FL}$ $10^8 \text{ s}^{-1}$ | $k_{rad}$ $10^8 \text{ s}^{-1}$ |
|---|---|---|---|---|---|---|---|---|
| TDI monomer | Toluene[1] | 652±1 | 663±1 | 3.4±0.1 | --- | 0.7 | 2.9±0.09 | 2.1 |
| TDI dimer | Toluene[1] | 665±2 | 675±2 | 2.2±0.04 | 80±17 | 0.64 | 4.5±0.08 | 2.9 |
| TDI Monomer | Zeonex[2] | --- | 650±3.5 | 3.6±0.25 | | --- | 2.8±0.2 | --- |
| TDI Dimer | Zeonex[2] | --- | 659±4 | 2.3±0.2 | 24.3±13.2 | --- | 4.3±0.4 | --- |

1: bulk solution measurement; 2: single molecule measurement

**Single Molecule Measurements**

**Measurements at room temperature.** Fig. 2 (a) and 2 (c) present the fluorescence intensity time-trace and spectrum of a representative TDI monomer molecule, respectively. The fluorescence spectra of single TDI molecules in zeonex are nearly identical to those observed in bulk toluene solution, with a distribution of peak positions (Supplementary Fig. 5 (a) and Table 1). The 13 nm spectral blue shift in the zeonex film compared to toluene is attributed to the different environments. Since the internal photophysical transitions control the sequence of photons emitted, the fluorescence intensity autocorrelation function $g^2(\tau)$ provides an ideal tool for measuring the rates of these transitions for immobilized single molecules[36,38–40]. In terms of photon counts or intensities, $g^2(\tau)$ is defined as[45,46]:

$$g^2(\tau) = \frac{\langle I(t)I(t+\tau)\rangle}{\langle I(t)\rangle^2} \qquad (2)$$

Exemplarily, $g^2(\tau)$ of a TDI monomer is shown in Fig. 2 (d) exhibiting the typical features resulting from the photon statistics of a single organic dye molecule.

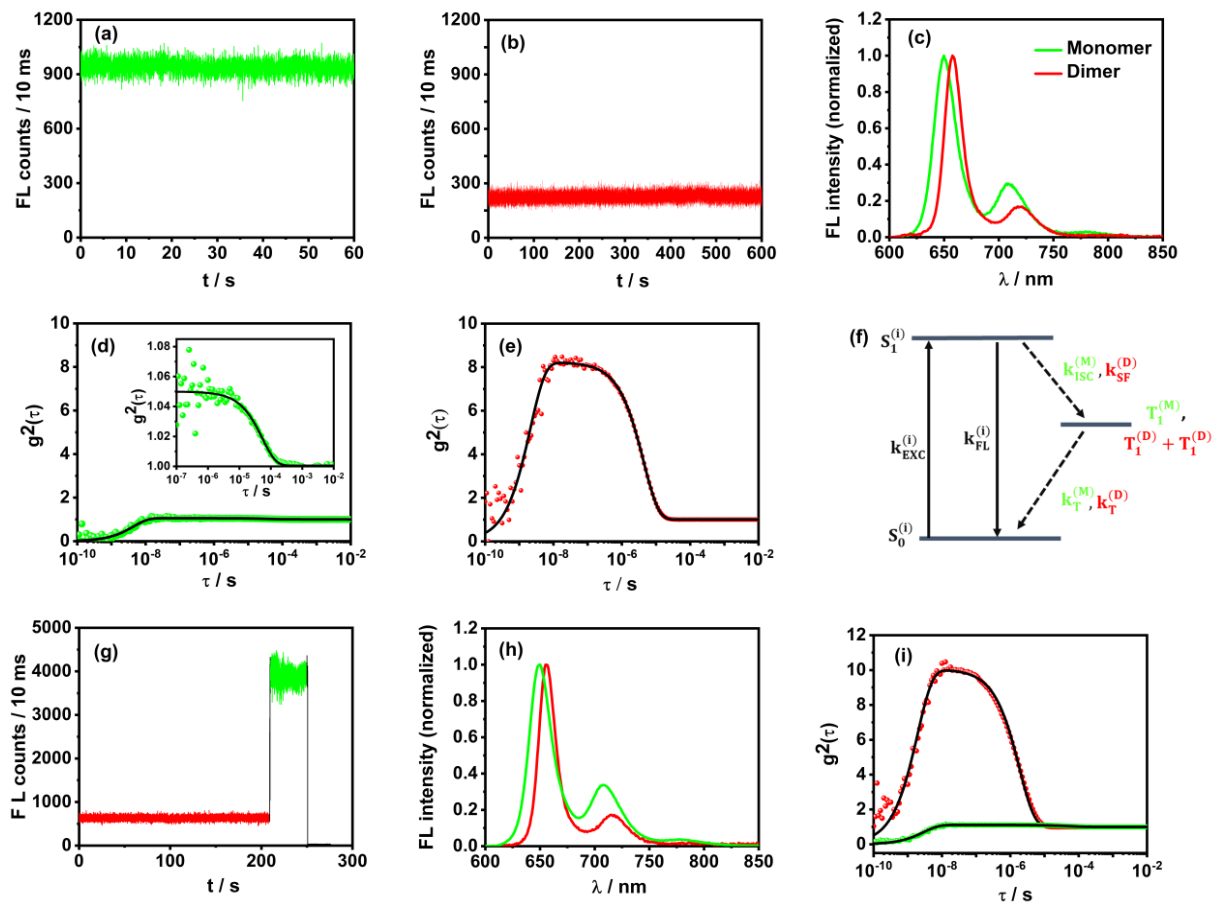

**Fig. 2 | ISC in individual TDI monomers vs. singlet fission in individual TDI dimers.** Fluorescence intensity time-traces of a single TDI monomer (a) and a single TDI dimer (b) and the corresponding fluorescence spectra (c). In accordance with the bulk solution data (Fig. 1b), the fluorescence spectrum of the dimer is red shifted with respect to the monomer and the intensity of the vibronic sideband is decreased. Fluorescence intensity autocorrelation functions $g^2(\tau)$ of a TDI monomer (d) and dimer (e). At short times, the correlation functions approach zero, indicating photon antibunching. At longer times (μs-ms), the decay of the correlation functions is due to photon bunching. In (f) a scheme for the photophysical transitions is presented to stress the analogy of transitions into a dark state in both the monomer (M) and dimer (D). Here $k_{exc}$, $k_{fl}$, $k_{ISC}$, $k_{SF}$ and $k_T$ represent excitation, fluorescence, intersystem crossing, singlet fission and triplet decay rates. Fluorescence time trace (g) of another single TDI dimer. After ~ 210 s a strong increase in the count rate is observed due to photobleaching of one of the monomers. The fluorescence spectra (h) and correlation functions (i) during the red and green phases. All data shown in this figure have been recorded using continuous-wave (cw) laser excitation.

To extract the transition rates for the TDI monomer from $g^2(\tau)$, we model it as an effective three-level system, following established methods in the literature[36–38,46]. In our experiments, the molecules are excited non-resonantly into a higher energy vibronic level of the $S_1$ state at room temperature. Under these conditions coherences can be ignored, simplifying the analysis to a system of rate equations that describe the transitions between the energy levels. Using this three-state model, $g^2(\tau)$ is given by the following expression[46]:

$$g^2(\tau) = -(1+C)e^{-\lambda_1 \tau} + (Ce^{-\lambda_2 t} + 1) \qquad (3)$$

Here, C, $\lambda_1$, and $\lambda_2$ represent the contrast, rise (antibunching component), and decay (bunching component) parameters of the correlation function, respectively. Since the typical excitation rates ($k_{exc} = 4 \cdot 10^5 s^{-1}$) were significantly lower than the fluorescence decay rate ($k_{fl} \sim 10^8 s^{-1}$), $\lambda_1 \approx k_{fl}$. The above parameters were then used as inputs to simultaneously calculate the ISC rate constant ($k_{ISC}$), the triplet decay rate ($k_T$) and the fluorescence lifetime ($\tau_{FL}$) in the TDI monomer through global analysis (see Methods for details). The data of the representative TDI monomer (Fig. 2 (d)) gave rise to an ISC rate of $k_{ISC} = 2.6 \cdot 10^5 s^{-1}$, a triplet decay rate of $k_T = 2.2 \cdot 10^4 s^{-1}$ and a fluorescence lifetime of $\tau_{FL} = 3.3$ ns. The distributions and average values of these parameters for TDI monomers are given in Supplementary Fig. 6. The fluorescence lifetime distribution peaks at 3.6 ns which is close to the value observed for TDI monomers in toluene solution (Table 1). Having the fluorescence decay rate and the triplet kinetics of the TDI monomer at our disposal, we have an ideal reference system to compare to the behaviour of the TDI dimer.

In Fig. 2 (b) and (c) the fluorescence time trace and spectrum of a single TDI dimer are shown. In accordance with the bulk solution data (Fig. 1 (b)), the fluorescence spectrum of the dimer (Fig. 2 (c) and Supplementary Fig. 5 (b)) is red shifted with respect to the monomer and the intensity of the vibronic sideband is decreased, corroborating the presence of the delocalized $S_1$ state in the dimer and directly signalling electronic coupling without reference to any photophysical rate parameters. The autocorrelation function of the dimer (Fig. 2 (e)) reveals a much higher contrast than that of the TDI monomer and the bunching part is shifted to shorter times. The correlation function could also be well fitted with Equation 3 (Fig. 2 (e)). Accordingly, we will provisionally approximate the dimer as an effective 3-level system undergoing transitions between optically bright and dark states. In line with the 3-level description of the monomer, we have defined the transition rates of the dimer as depicted in the scheme in Fig 2 (f). $k_{SF}$ denotes the overall formation rate of the SF-born triplets. This process may involve several intermediate states to be discussed later. The decoupled triplets then decay with the rate $k_T$. Solving again for the 3-level system population dynamics, we find for the dimer in Fig. 2 (e), $k_{SF} = 2.9 \cdot 10^8 \text{ s}^{-1}$, $k_T = 2.7 \cdot 10^4 \text{ s}^{-1}$ and a fluorescence lifetime of 2.1 ns.

The characteristic differences between TDI dimers and monomers are nicely reproduced in the sequence of events once one of the monomers bleaches in a TDI dimer (Fig. 2 (g)). In the fluorescence time trace (Fig. 2 (g)), the initial fluorescence count rate is low (red phase) and switches suddenly to a much higher level at ~ 210 s (green phase). The count rate has increased appreciably and remains stable until the signal irreversibly drops to the background level. The fluorescence spectrum (Fig. 2 (h)) during the green phase is blue shifted (compared to the red phase) and shows an increased intensity of the vibronic side band (0-1 transition). These observations indicate that during the red phase an intact TDI dimer is active, while the data in the green phase originate from a TDI monomer. Accordingly, at the end of the red phase one of the TDI monomers has selectively photobleached which is a quite common event when studying individual multichromophoric compounds[47]. The correlation functions in the two phases (Fig. 2 (i)) show the characteristic differences in correlation contrast to be expected for the dimer (red) and monomer (green), respectively. In the red phase, $k_{SF} = 3.5 \cdot 10^8 \text{ s}^{-1}$ is found being significantly higher than the ISC rate $k_{ISC} = 8.9 \cdot 10^5 \text{ s}^{-1}$ obtained for the monomer in the green phase. The $k_T$ values for the red and green phase are $5.2 \cdot 10^4 \text{ s}^{-1}$ and $1.6 \cdot 10^4 \text{ s}^{-1}$, respectively. The fluorescence lifetimes in the two phases are 1.9 ns (red) and 2.9 ns (green), again substantiating the assignment to dimer and monomer emission. This behaviour was consistently observed for many TDI dimers.

As is evident from the fluorescence intensity time-traces (Fig. 2(a), (b) and (g)), TDI dimers exhibit significantly lower count rates than monomers. Since the fluorescence lifetimes remain unchanged from toluene solution to zeonex, the quantum yields at the bulk and single molecule level are expected to be similar in both matrices. Accordingly, the smaller count rates in the dimers do not reflect a decrease of the fluorescence quantum yield but are primarily due to early saturation of the emission rate because of efficient population of the triplet bottleneck.

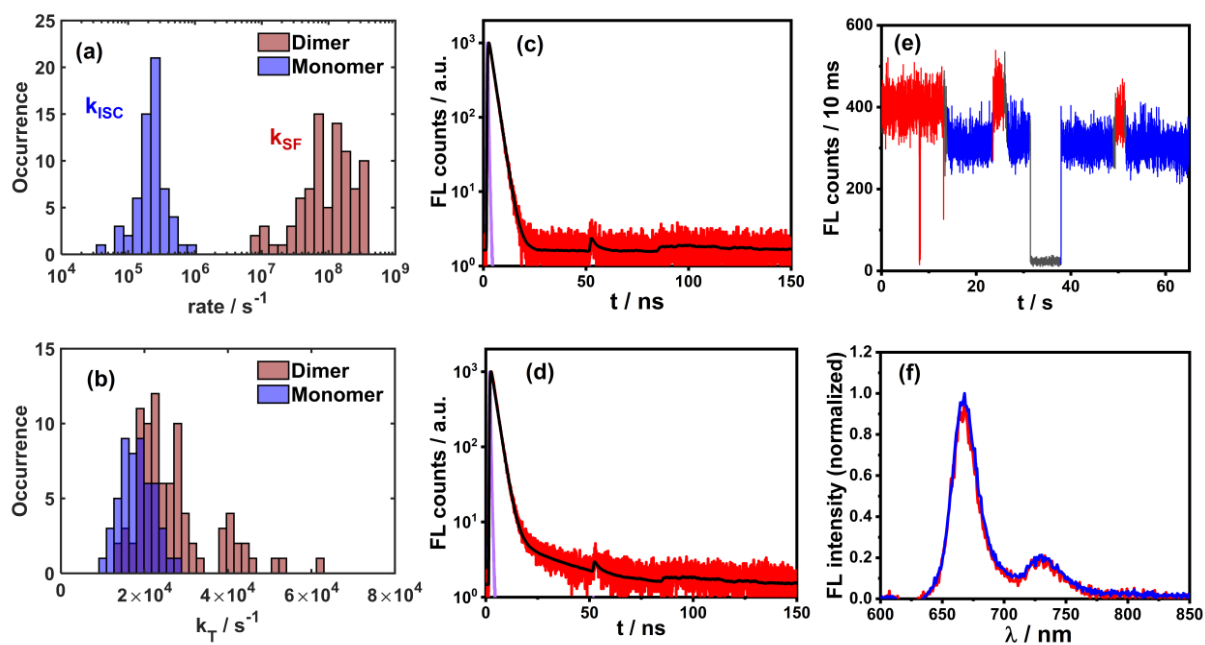

**Fig. 3 | Triplet kinetics, fluorescence decays and count rate fluctuations reveal static and dynamic heterogeneities.** (a) Distributions of $k_{SF}$ and $k_{ISC}$ for TDI dimers (brown) and monomers (blue). (b) Distributions of $k_T$ for TDI dimers (brown) and monomers (blue). The distributions are derived from data recorded under both cw and pulsed excitation. Fluorescence decay curves of two single TDI dimers in zeonex (c, d). The dimer in (c) gave rise to a single exponential FL decay (prompt fluorescence) with a lifetime of 2.4 ns. In contrast, the dimer in (d) showed a bi-exponential decay with contributions from prompt fluorescence ($\tau_{FL} = 2.4$ ns, 94%) and delayed fluorescence ($\tau_{del} = 24.5$ ns, 6%). Fluorescence time trace (e) of a single TDI dimer exhibiting slow fluctuations of the count rate between two levels (red and blue). Emission spectra (f) recorded during the red and blue phases. No differences in peak position or spectral shape were observed, demonstrating that the electronic coupling remained unchanged throughout the entire sequence.

The distributions of $k_{SF}$ and $k_{ISC}$ for TDI dimers and monomers, respectively, are presented in Fig. 3 (a). The average value of $k_{SF}^{avg} = 1.3 \cdot 10^8 \, s^{-1}$ is more than 500 times larger than the average ISC rate in the monomer indicating efficient population of triplet states. When comparing $k_T$ in the dimer and monomer (Fig. 3 (b)), it is seen that the bulk of the dimer distribution largely overlaps the monomer distribution. $k_T^{avg} = 2 \cdot 10^4 \text{ s}^{-1}$. The dimers with larger $k_T$ values may represent cases where the decay is accelerated by triplet-triplet annihilation assuming that preferentially two triplet states ($T_1 + T_1$) are generated[21].

So far, the discussion has referred to data obtained with continuous-wave (cw) excitation. In addition, experiments with pulsed excitation were performed. In Fig. 3 (c), (d), the fluorescence decay curves of two individual TDI dimers are displayed. In one case (Fig. 3 (c)) a single exponential decay was observed with a fluorescence lifetime of 2.4 ns, while in the other case (Fig. 3 (d)) the decay was biexponential which we assign to prompt $(\tau_{FL} = 2.4 \text{ ns } (94\%))$ and delayed fluorescence ($\tau_{del} = 24.5$ ns (6%)). Such biexponential decays were observed for 20 % of the dimers. The single molecule delayed fluorescence lifetimes appeared to be shorter than the bulk value in toluene and were distributed between 10 ns and 40 ns (see Supplementary Fig. 7). We note, however, that due to signal-to-noise limitations in the single molecule measurements decays with longer lifetimes may have escaped detection. The data indicate that the occurrence of delayed fluorescence and its contribution to the total decay strongly varies from dimer to dimer. Considering that delayed fluorescence is a thermally activated process, the interaction of a given dimer with its particular local environment will give rise to a wide distribution of barrier heights and rates. This view is corroborated by the low temperature data.

In Supplementary Fig. 8 the distributions of prompt fluorescence lifetimes of single TDI dimers obtained from the correlation analysis after cw excitation and from the fluorescence decay curves after pulsed excitation are presented. The average values (2.2 ns and 2.3 ns, respectively) agree within experimental error and match the bulk value (2.2 ns), confirming the reliability of the photophysical parameters derived from the multi-parameter correlation analysis.

The distributions presented in Fig. 3 (a.b) as well as the observation of varying fluorescence decay behaviour have shown that the photophysics varies appreciably from dimer to dimer. In addition to these static heterogeneities, we have observed dynamic heterogeneities referring to changes in the behaviour of a given dimer in the course of time. In the example shown in Fig. 3 (e), the fluorescence count rate is fluctuating on a slow time scale between two levels. The reversible transition to a dark state at ~ 30 s will not be considered further. The emission spectra do not show any differences in position and shape between the red and blue phases (Fig. 3 (f)) clearly demonstrating that the electronic coupling does not change throughout the whole sequence. During the red as well as the blue phases the fluorescence decays are biexponential revealing prompt and delayed fluorescence. The prompt fluorescence lifetime – dominated by the radiative rate – remains constant during the whole time-trace ($\tau_{FL}$ = 2.2 ± 0.04 ns). In contrast, the lifetime of the delayed fluorescence is different in the red and blue phases. In the red phase the average value is $\tau_{del}$ = 12 ± 1.2 ns while in the blue phase we find $\tau_{del}$ = 17 ± 2.8 ns. For later comparison, we convert the delayed fluorescence lifetimes into rates (red phase: $k_{del}$ = 8.2 ± 0.8 · $10^7$ $s^{-1}$; blue phase: $k_{del}$ = 5.9 ± 1.0 · $10^7$ $s^{-1}$). In addition, $k_{SF}$ has been determined from the correlation function for both phases (red phase: $k_{SF}$ = 4.2 ± 0.4 · $10^7$ $s^{-1}$; blue phase: $k_{SF}$ = 6.8 ± 0.7 · $10^7$ $s^{-1}$). When comparing $k_{del}$ in the red and blue phase, it is seen that it is smaller in the blue phase. In contrast, $k_{SF}$ is larger in the blue and smaller in the red phase. Within the error margins the sum of both rates ($k_{del} + k_{SF}$) in each of the two phases is constant indicating that the two rates are not changing independently. Moreover, with an increased (non-radiative) rate $k_{SF}$ in the blue phases, the fluorescence count rate roughly drops by the same factor (Fig. 3 (e)).

In Supplementary Fig. 9, distributions and average values of $k_{SF}$ and $k_T$ are presented for dimers which exhibited either only prompt fluorescence or both prompt and delayed fluorescence.

When delayed fluorescence was observed, the average $k_{SF}$ value was smaller than for dimers which showed only prompt fluorescence. In contrast, the average $k_T$ value did not depend on the absence or presence of delayed fluorescence. Interestingly, the photostability of dimers which showed delayed fluorescence was larger than for those with prompt fluorescence only. Apparently, a more efficient back transfer into the $^1(S_1S_0)$-state protects the molecules from photobleaching which is thought to occur more efficiently from the triplet state[48].

**Measurements at cryogenic temperature (T = 1.4 K).** To address the notion of a coherent superposition state, we also performed measurements at 1.4 K at which the effects of thermal broadening and activation are largely suppressed. The fluorescence lifetime at 1.4 K (Fig. 4 (a)) was basically the same as at room temperature, but no delayed fluorescence was observed. In the low temperature measurements, however, a reliable analysis of the intensity correlation function was prevented, because the excitation rate could not be determined accurately. Since the fluorescence lifetime did not change, we assume that the singlet fission rate ($k_{SF}$) also remained unaffected.

Furthermore, we have measured fluorescence emission and excitation spectra at 1.4 K of monomers and dimers. As seen in Supplementary Fig. 10, the monomers exhibit a sharp and intense [0,0] zero-phonon line (ZPL) and several well-resolved vibronic transitions. In the dimers a low-frequency mode at 80 cm$^{-1}$ dominates the vibronic structure. A comparison of the monomer and dimer spectra in the low-frequency region (Fig. 4 (b)) highlights the emergence of this intermolecular mode, previously observed by femtosecond stimulated Raman spectroscopy (FSRS) spectroscopy and tentatively assigned to a breathing mode with torsional contributions[31].

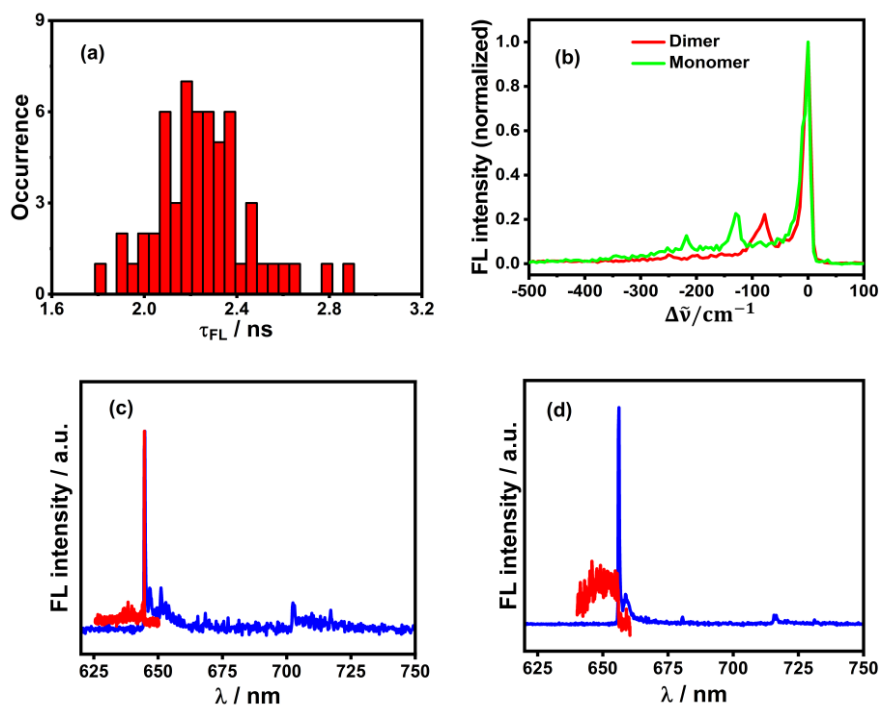

**Fig. 4| Singlet fission and electron-phonon coupling at T = 1.4 K.** (a) Fluorescence lifetime ($\tau_{FL}$) distribution of the TDI dimer derived from fluorescence decay measurements at 1.4 K. The average lifetime calculated from the distribution ($\tau_{FL}^{avg} = 2.3 \pm 0.2$ ns) is identical to the bulk and single molecule values obtained at room temperature. (b) Representative fluorescence spectra of a single TDI monomer (green) and dimer (red). Fluorescence emission (blue) and excitation (red) spectra of a TDI monomer (c) and a dimer (d).

In Fig. 4 (c) the fluorescence emission and excitation spectra of a TDI monomer are displayed. In both types of spectra, resonant [0,0]-ZPLs are observed. The observed linewidth of the ZPL in the excitation spectrum is limited by the laser bandwidth (3 GHz), while the actual linewidth should be close to the lifetime limit (50 MHz). In contrast, TDI dimers exhibit a markedly different behaviour. While the emission spectrum still showed a sharp ZPL and the characteristic 80 cm$^{-1}$ mode, the excitation spectrum typically consisted of only a broad band (Fig. 4 (d)). In rare cases, a very weak ZPL could be discerned at the low energy edge of the broad band. For an analogous N-N-coupled collinear Perylenediimid-dimer, in which no SF occurs, [0,0]-ZPLs were observed in absorption and emission[43]. The lack of symmetry between excitation and emission spectra in the TDI dimer clearly indicates that the absorbing and emitting states are different. The absence of a ZPL in the excitation spectrum is a consequence of strong electron-phonon coupling which appears to be characteristic of the initially photoexcited state. In agreement with previous suggestions[30,31], we assume that this state is a coherent ($^1(S_1S_0) \leftrightarrow (T_1T_1)$) superposition. In addition, our data let us to conclude that $^1(S_1S_0)$ evolves from the decaying superposition state as outlined in the next section.

**Discussion**

Considering our single molecule and pertinent literature data, we propose a simplified scheme in Fig. 5 for the formation and decay of SF born triplet states. This model assumes the initial formation and decay of a coherent $^1(S_1S_0) \leftrightarrow {}^1(T_1T_1)$ superposition state strongly supported by our low temperature single molecule measurements. As a next step, the superposition state decays into either directly into $^1(T_1T_1)$ or into $^1(S_1S_0)$. A similar scheme has been suggested very recently for SF in a rubrene crystal[49]. The high fluorescence quantum yield of the dimer suggests that the superposition state decays predominantly into $^1(S_1S_0)$. Regarding the formation of the pure $^1(S_1S_0)$ state, we emphasize that the data for this delocalized singlet state did not show any admixture of the $^1(T_1T_1)$ state. The pure $^1(S_1S_0)$ state either decays to $S_0$ or $^1(T_1T_1)$ is formed. Next, the spin pure $^1(T_1T_1)$ state can develop into an intermediate mixed spin state $^m(T_1 \cdots T_1)$ which in turn either leads to delayed fluorescence (see below) or to independent triplets ($T_1 + T_1$) eventually decaying to $S_0$. In the single molecule correlation measurements, however, the different dark triplet states depicted in Fig. 5 appear as a single dark state and cannot be distinguished. Consequently, the rate constant $k_{SF}$, provisionally defined in Fig. 2 (f), has to be understood as the overall rate-determining SF step, encompassing all processes that convert the fluorescent $^1(S_1S_0)$ state into uncoupled triplets ($T_1 + T_1$). The average $k_{SF}$ value of $1.3 \cdot 10^8$ s$^{-1}$ for TDI dimers in zeonex is close to the formation rate of uncorrelated triplets (8 · $10^7$ s$^{-1}$) deduced from the modelling of nsTA data for the TDI dimer in chlorobenzene[30]. Notably, the latter value is almost identical to the value obtained from our bulk solution data.

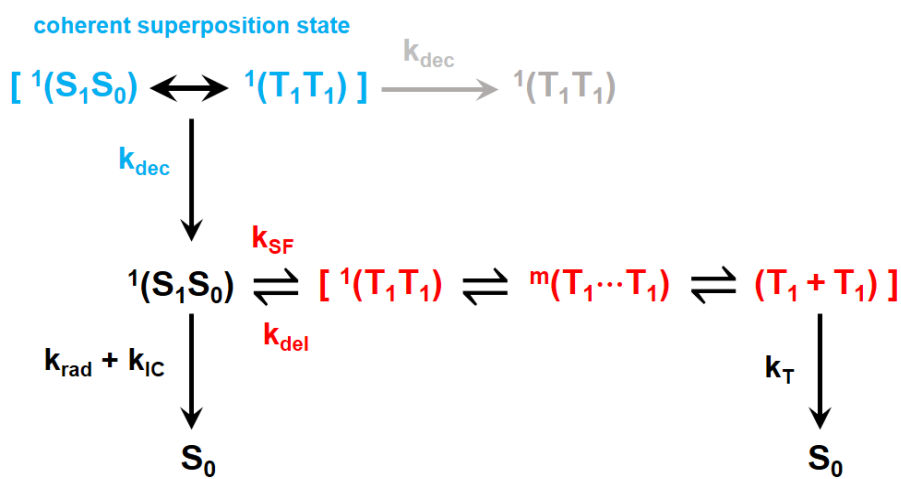

**Fig. 5 | Scheme to describe photophysical transitions and spin evolution in the TDI dimer.** The model assumes the initial formation of a coherent superposition state, $^1(S_1S_0) \leftrightarrow {}^1(T_1T_1)$. As a next step, the superposition state decays into either $^1(S_1S_0)$ or directly into $^1(T_1T_1)$. The corresponding rates $k_{dec}$ are not accessible in our experiments. From the pure $^1(S_1S_0)$ state, either the $^1(T_1T_1)$ or the singlet ground state ($S_0$) are populated. Next, the spin-pure $^1(T_1T_1)$ state may evolve into an intermediate mixed-spin state $^m(T_1\cdots T_1)$. Since the various dark state transitions in the triplet manifold may have an impact on the rates of SF and delayed fluorescence, these rates may reflect the sum of several processes. The independent triplets ($T_1 + T_1$) eventually decay to $S_0$.

Time resolved EPR spectroscopy had disclosed that the spin multiexciton state $^m(T_1 \cdots T_1)$ of the TDI dimer is composed of a mixture of singlet and quintet states and that without the presence of a magnetic field the formation of a $^3(T_1 \cdots T_1)$ state can be neglected and almost all triplet formation accounted for by triplet dissociation[30]. Accordingly, out of the mixed $^m(T_1 \cdots T_1)$ spin state uncoupled triplets are formed. Since we have found the distribution of triplet decay rates $k_T$ to be bimodal (Fig. 3 (b)), the fraction with a faster decay than observed for the TDI monomer may reflect triplet-triplet annihilation.

In covalently coupled, spatially isolated dimers the triplets cannot diffuse away. Therefore, spin dephasing might be prevented, in which case the triplets preferentially annihilate back to the singlet ground state $S_0$[18,21,22,34]. In our case, however, we clearly have observed a correlation decay on a time scale of the triplet lifetime. Generally, it is thought that structural changes are needed to induce spin dephasing. In the TDI dimer, the monomers are twisted against each other by almost 90° which leads to small orbital overlap and weak exchange coupling. In our low temperature measurements, a prominent low frequency intermolecular mode at 80 cm$^{-1}$ has been observed. We assume that such a mode can modulate the overlap and exchange coupling between the two TDI units[50]. This appears to be a route to induce spin dephasing and generate uncoupled triplets.

Besides dissociating into uncoupled triplets, the mixed spin state can revert back to the $^1(S_1S_0)$ state by triplet fusion which leads to delayed fluorescence. Delayed fluorescence was observed for approximately 20% of the dimers at room temperature but disappeared completely at low temperature, indicating a potential barrier to triplet fusion that depends on the local environment and cannot be overcome at 1.4 K. Now, the question arises what causes the correlated variations of the rates $k_{SF}$ and $k_{del}$ in the course of time, as shown in Fig. 3 (e). We assume that the coupling strength is related to the dihedral angle between the two TDI units. Prolonged irradiation of a single molecule and concomitant heat dissipation in its local environment can lead to a conformational change by which the dihedral angle between the two TDI units changes temporarily. In this scenario, the small orbital overlap between the two π-systems is altered leading to a variation of the composition of the mixed spin state. Because the probability of delayed fluorescence scales with the singlet content of the wavefunction, an increased singlet contribution enhances triplet fusion ($k_{del}$), whereas greater quintet character promotes triplet dissociation ($k_{SF}$) [18]. Throughout the whole sequence depicted in Fig. 3 (e), the uniform fluorescence lifetime and spectra did prove that electronic coupling in the dimer remained essentially constant. This behaviour is consistent with our model, because a change of the dihedral angle will not impact the coupling strength in $^1(S_1S_0)$ considering that the transition dipole moment of TDI is oriented along the long molecular axis[43].

In conclusion, SF between two single molecules has been reported for the first time. The single molecule approach has led to a number of novel insights which would be difficult to obtain at the bulk level. The finding of static and dynamic heterogeneities sheds light on the origin of the omnipresent complexity of the SF process. Fluorescence intensity fluctuations of single dimers allow for accessing spin evolution by purely optical means. Experiments at 1.4 K – rarely reported in the SF field – support the initial formation of a superposition state as well as the multistep nature of the SF process. Overall, this study opens new avenues for further exploration of SF in other organic chromophores at the single-molecule level and may also provide valuable insights into the photophysics of thermally activated delayed fluorescence molecules, which are known for their very large ISC rates[51].

**Methods**

**Bulk Absorption and Fluorescence Spectroscopy**

Absorption and emission spectra were recorded using a Duetta absorbance and fluorescence spectrometer (Jobin-Yvon) with a sensitivity extending up to 1100 nm. Fluorescence decays were acquired by time-correlated single photon counting (TCSPC) measurements on a modified Fluorolog-3 spectrofluorometer (Jobin-Yvon). The setup was equipped with a pulsed fibre laser (YSL SC-OEM + YSL VLF) and a hybrid detector (PicoQuant, PMA Hybrid 50), both connected to a TCSPC module (PicoQuant, PicoHarp 300) yielding a time resolution of ~300 ps.

**Calculation of the electron coupling strength and coherence number**

Assuming through space electrostatic dipole-dipole coupling, the coupling strength V in the TDI dimer was estimated by equation 4:

$$V = \frac{1}{4\pi\varepsilon_0} \frac{|\mu^2|\kappa}{R^3} \qquad (4)$$

where μ = 11 Debye is the transition dipole moment of the TDI monomer in toluene[52], κ = 2 the orientational factor and R = 1.73 nm the centre-to-centre distance in the dimer. V was found to be 240 cm$^{-1}$.

It has been shown by Spano et al.[42] that for J-type aggregates with a dominating vibrational mode and in the presence of disorder and T > 0 K, the coherence number N$_{coh}$ can be closely approximated by equation 5:

$$\frac{I_{0,0}}{I_{0,1}} = \frac{N_{coh}}{S} \qquad (5)$$

$I_{0,0}$ and $I_{0,1}$ are the integrated intensities of the 0,0-transition and 0,1-transition in the fluorescence spectrum and S is the Huang-Rhys factor of the coupled vibrational mode. From the fluorescence spectrum of the TDI dimer in toluene we obtain $I_{0,0}/I_{0,1}$ = 1.9 and from the TDI monomer fluorescence spectrum S = 0.74, a typical value for a rigid organic dye molecule. Putting these numbers into equation 5, we get N$_{coh}$ = 1.4. Qualitatively, this number is related to the degree of delocalization of the electronic excitation in the TDI dimer in the presence of disorder and at finite temperature. In terms of radiative rates, N$_{coh}$ is given by the ratio of the dimer to monomer radiative rates. Using this relation, we find a radiative rate in the TDI dimer of $k_{rad}^{Dimer} = k_{rad}^{Monomer} \cdot 1.4 = 2.9 \cdot 10^8 \text{ s}^{-1}$ which is in full agreement with the radiative rate obtained from the quantum yield and lifetime.

**Single Molecule Spectroscopy**

Thin-film samples for single molecule measurements were obtained by spin-coating toluene/zeonex 330R solutions containing TDI/TDI dimer at concentrations ranging from 10$^{-10}$ M to 10$^{-11}$ M on top of cleaned glass cover slides. Zeonex a non-polar hydrocarbon polymer was chosen because its low dielectric constant (e = 2.5) which is close to that of toluene and should prevent contributions of charge transfer states to SF as was reported in the literature[29]. Single molecule measurements at room temperature were conducted with a home-built confocal fluorescence microscope. Excitation was provided either by a fibre-coupled cw solid-state laser (Coherent OBIS, 594 nm) or, in the case of pulsed operation, by a pulsed white-light fiber laser (SC-OEM, YSL Photonics, China) with a variable optical filter (YSL VLF) tuned to ~594 nm. The laser beam was collimated and focused onto the sample plane using an oil-immersion objective (Zeiss, Plan-Apochromat 63×, NA = 1.4) after passing through a 594 nm band-pass filter and reflecting off an 80T:20R beam splitter. The laser power at the sample was ~5 µW, corresponding to an excitation intensity of ~1 kW/cm², with argon gas flowing continuously to minimize photobleaching of TDI monomers and dimers. Fluorescence was collected through the same objective, with the excitation light blocked by a 615 nm long-pass filter in the detection path. The fluorescence light was then split into two paths by a 50T:50R beam splitter. One path was further divided and sent to two avalanche photodiodes (APDs) in a Hanbury-Brown Twiss configuration to detect photon arrival times, which were recorded using a HydraHarp 400 TCSPC module (PicoQuant). The second path was directed to a spectrograph (Acton Spectra Pro 300i, resolution ~25 cm$^{-1}$) equipped with EM-CCD camera (ProEM HS, Teledyne Princeton Instruments).

Measurements at 1.4 K were performed using a home-built variable-temperature confocal microscope setup. To record fluorescence emission spectra, excitation was provided by the same cw or pulsed lasers used in the room-temperature experiments. The excitation light was focused onto the sample with a microscope objective (Melles Griot, 60×, NA 0.85) mounted inside an optical cryostat and immersed in liquid helium. Fluorescence emission was collected by the same objective, filtered with a 615 nm long-pass filter, and split by a 50T:50R beamsplitter. One fraction was directed to two APDs in a Hanbury–Brown–Twiss configuration as described above. The other fraction was dispersed by a spectrograph (Acton Spectra Pro 500i, resolution ~20 cm$^{-1}$) and detected with an EM-CCD camera (Newton, Andor). Fluorescence excitation spectra were acquired using a tuneable ring dye laser (Coherent 899-01) operated in broadband mode (3 GHz bandwidth) with a DCM dye solution providing a scan range of 620 – 670 nm. Red-shifted fluorescence was collected using a 685 nm long-pass filter and detected by the two APDs. Excitation intensities during the scans ranged from 3 to 152 W/cm².

**Fluorescence Intensity Autocorrelation Analysis and Estimation of the rates $k_{ISC}$, $k_{SF}$ and $k_T$**

Fluorescence intensity autocorrelation functions were calculated from the photon arrival times recorded with the HydraHarp 400. Generation and fitting of the correlation functions were performed using home-written MATLAB routines. For the full theoretical treatment of the model, including the derivation of the rate constants, we refer to the literature[46]. In our analysis, the triplet population rate is denoted as $k_{ISC}$ for the TDI monomer and $k_{SF}$ for the dimer, reflecting their different physical origins (intersystem crossing vs. singlet fission). Mathematically, both processes are represented by the same parameter in the fitting model.

The fitting parameters $\lambda_1$, $\lambda_2$, and C obtained from Equation (3) are given by:

$$\lambda_1 = \frac{1}{2}\left(a + d - \sqrt{(a-d)^2 + 4bc}\right) \quad (6)$$

$$\lambda_2 = \frac{1}{2}\left(a + d + \sqrt{(a-d)^2 + 4bc}\right) \quad (7)$$

$$C = -\frac{\lambda_1(1 + \lambda_2/f)}{(\lambda_1 - \lambda_2)} \quad (8)$$

a, b, c, d and f are given by the following expressions:

$$a = -(k_{exc} + k_T),\ b = \left(\frac{1}{\tau_{FL}} - k_{\frac{ISC}{SF}} - k_T\right),\ c = k_{exc},\ d = -\frac{1}{\tau_{fl}},\ \text{and}\ f = k_T$$

**Acknowledgements**

This work was supported by the Deutsche Forschungsgemeinschaft (DFG), Project No. 429529648, TRR 306 QuCoLiMa (Quantum Cooperativity of Light and Matter).

# Supplementary Information

**S1. Synthesis of TDI dimer**

The synthesis of the TDI dimer was achieved through a two-step process, following the same procedure previously reported by Wasielewski and coworkers[1]. In the first step, Suzuki coupling between compounds 1 and 2 yielded compound 3 with a 70% yield. This was followed by a base-induced dehydrogenation reaction using $K_2CO_3$, resulting in the formation of the fully conjugated TDI dimer skeleton. The crude product was purified by column chromatography on $SiO_2$ ($CH_2Cl_2$ / methanol=50:1) followed by size exclusion chromatography (Bio-Rad Bio-Beads, S-X1, THF). $^1$H NMR (500 MHz, $C_2D_2Cl_4$, 403K) δ 8.65 (8H, br, ArH), 8.51 (16H, br, ArH), 5.20 (2H, br, ArH), 2.29 (8H, m, $CH_2$), 1.97 (8H, m, $CH_2$), 1.33 (30H, m, $CH_2$), 0.90 (12H, t, J= 6.5 Hz, $CH_3$). HRMS MALDI-TOF (TCNQ): calculated for $C_{98}H_{84}N_4O_8$ m/z = 1446.6446 found m/z = 1446.6385. MP > 400 °C.

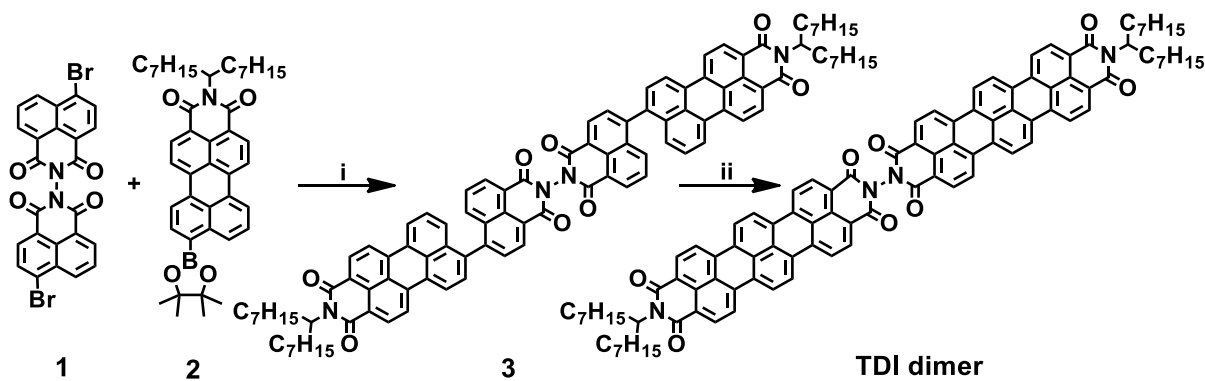

**Supplementary Fig. 1|** Synthesis of the TDI dimer. Reaction conditions: (i) Tetrakis (triphenylphosphine) palladium (0), $K_2CO_3$, toluene, 100°C, 12 h, 70%. (ii) $K_2CO_3$, ethanolamine, 160°C, 12 h, 10%.

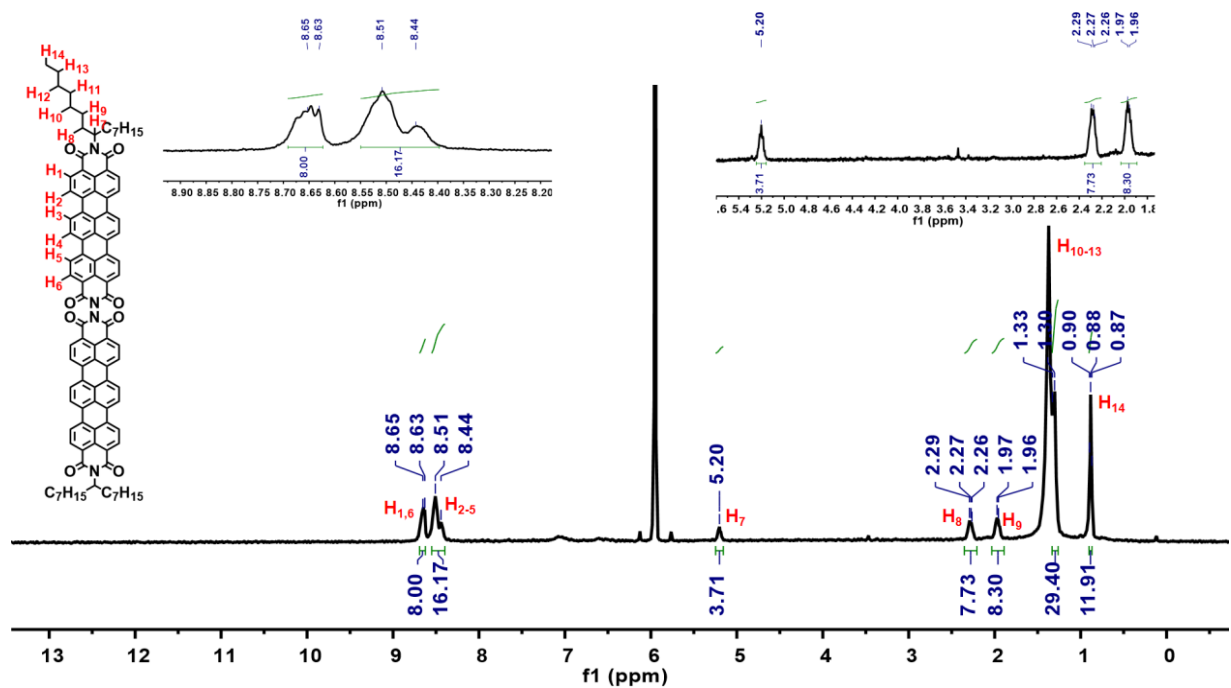

**Supplementary Fig. 2**| ¹H NMR spectrum of TDI dimer (500 MHz, $C_2D_2Cl_4$, 403K).

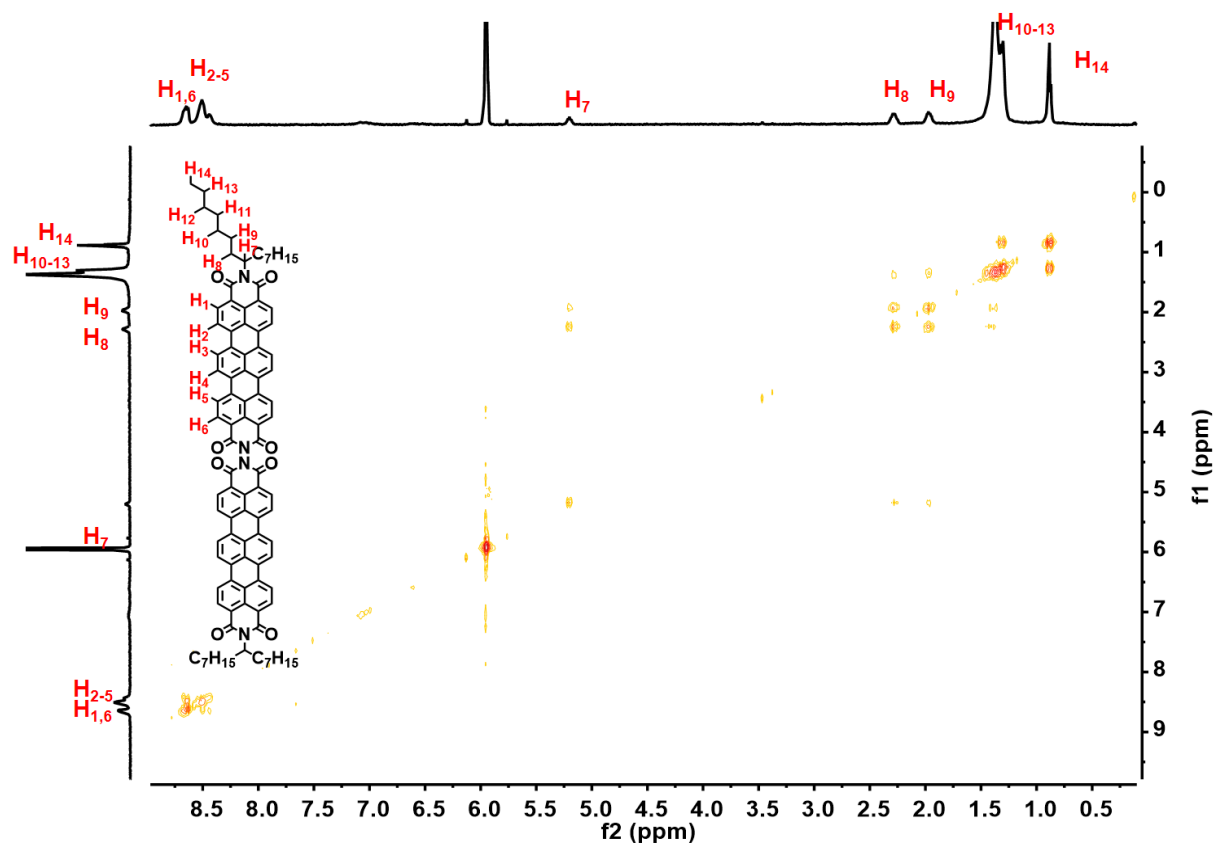

**Supplementary Fig. 3**| ¹H-¹H COSY spectrum of TDI dimer (500 MHz, $C_2D_2Cl_4$, 403K).

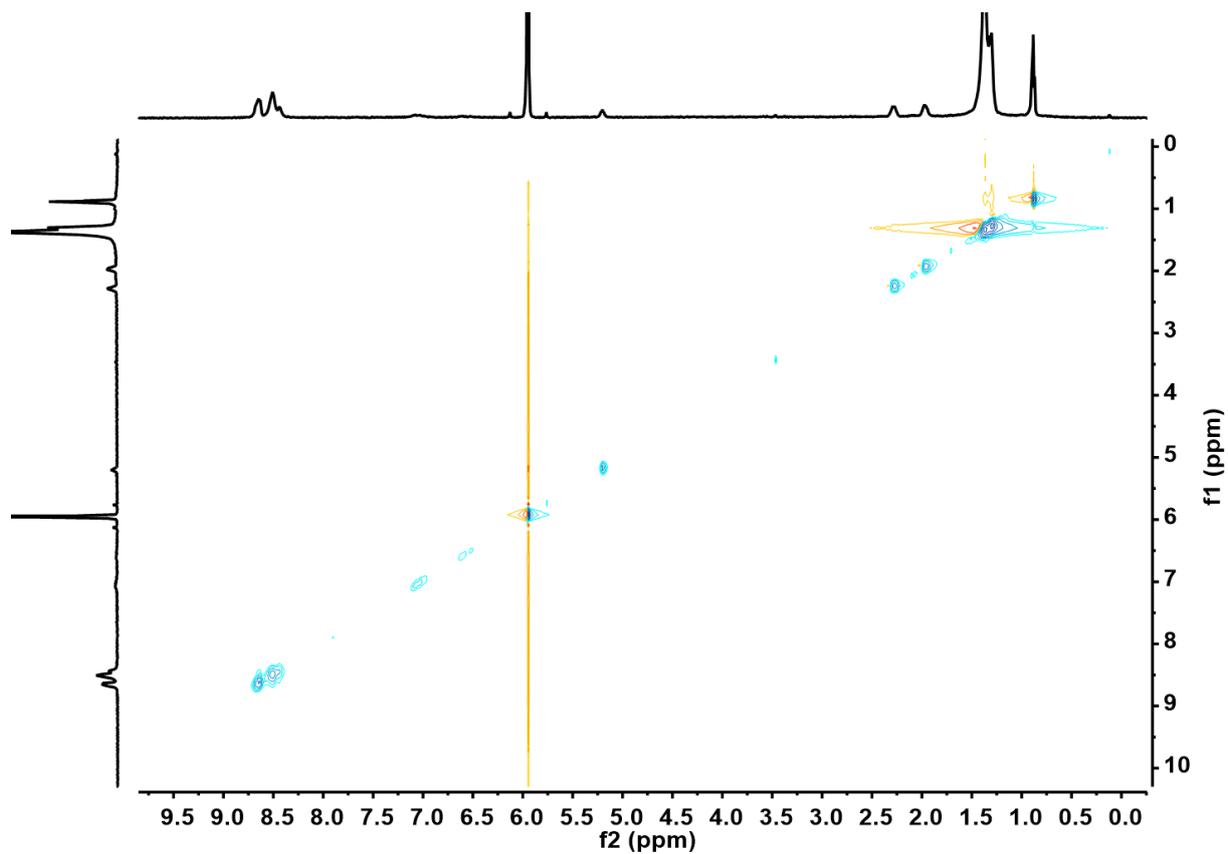

**Supplementary Fig. 4**| ¹H-¹H NOESY spectrum of the TDI dimer (500 MHz, C$_2$D$_2$Cl$_4$, 403K).

## S3. Single molecule distributions of spectral and photophysical parameters at room temperature

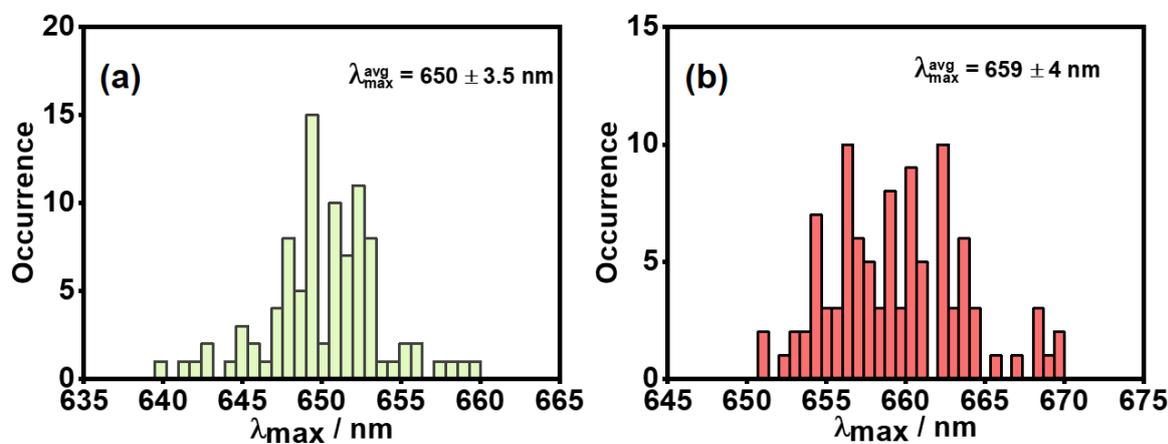

**Supplementary Fig. 5|** Distribution and average values of fluorescence peak positions of individual TDI monomers (a) and dimers (b).

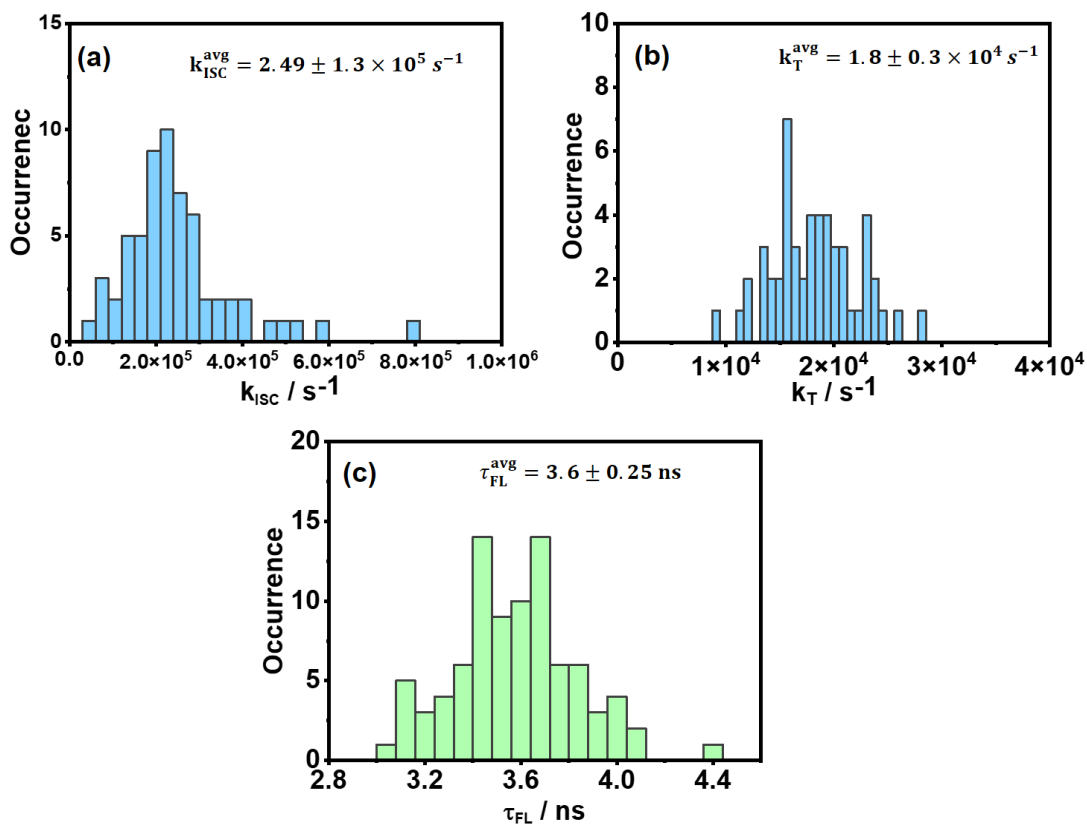

**Supplementary Fig. 6|** Distributions and average values of the ISC rate (a), triplet decay rate (b) and fluorescence lifetime (c) of the TDI monomer.

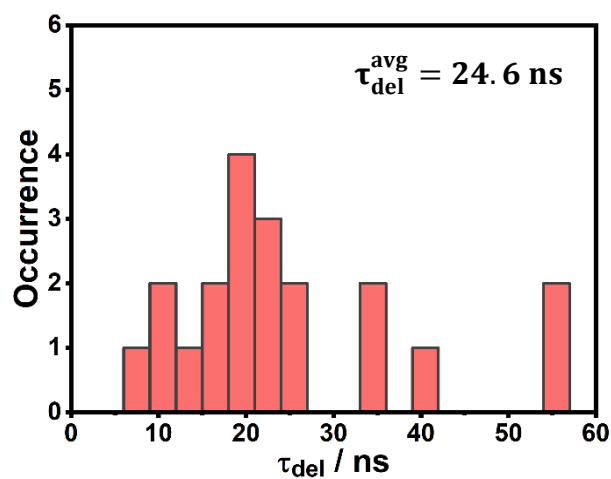

**Supplementary Fig. 7|** Distribution of delayed fluorescence lifetime for the TDI dimer after pulsed excitation.

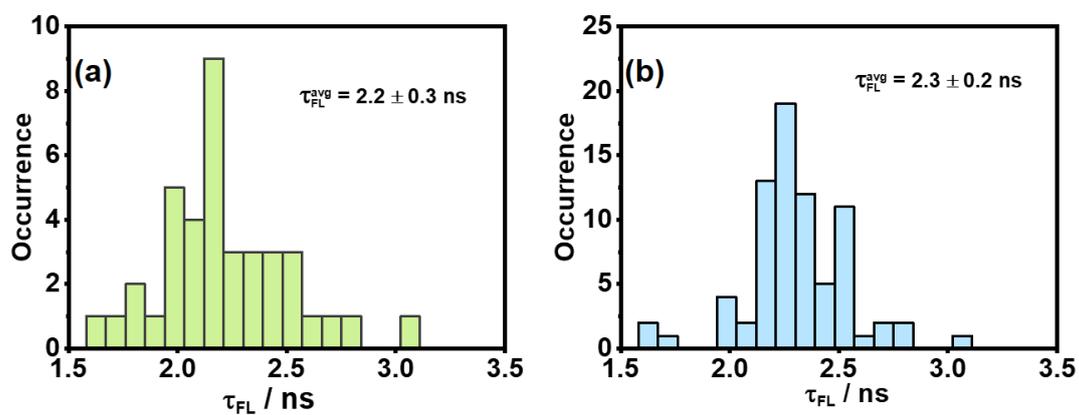

**Supplementary Fig. 8|** Distributions of the fluorescence lifetime ($\tau_{FL}$) of the TDI dimer derived from autocorrelation analysis after continuous wave excitation (a) or from fluorescence decay measurements after pulsed excitation (b).

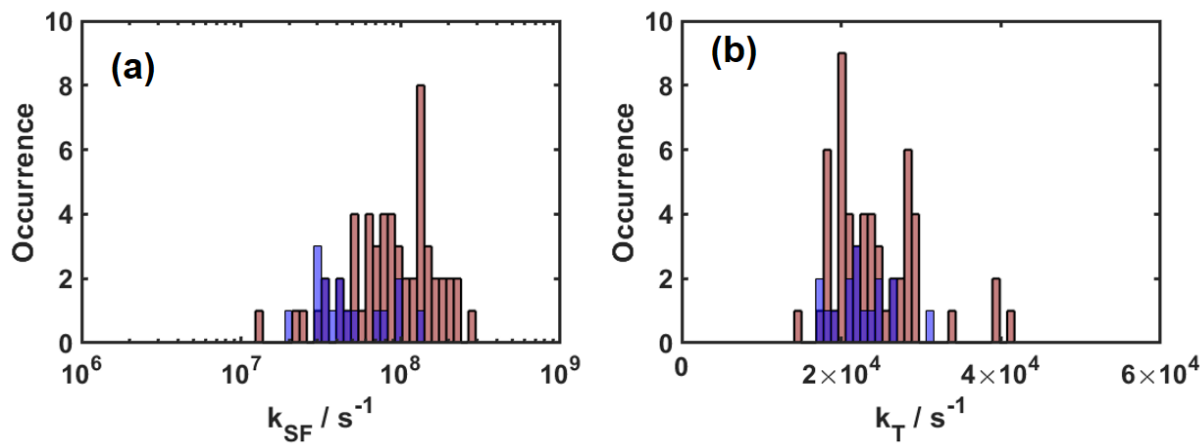

**Supplementary Fig. 9|** Distributions of the rates $k_{SF}$ (a) and $k_T$ (b) for TDI dimers which showed only prompt fluorescence (marrone) or prompt as well as delayed fluorescence (blue). The average values are: Prompt fluorescence: $k_{SF} = 1.03 \pm 0.6 \cdot 10^8$ s$^{-1}$, $k_T = 2.4 \pm 0.6 \cdot 10^4$ s$^{-1}$ ; prompt and delayed fluorescence: $k_{SF} = 5.5 \pm 3 \cdot 10^7$ s$^{-1}$, $k_T = 2.2 \pm 0.4 \cdot 10^4$ s$^{-1}$.

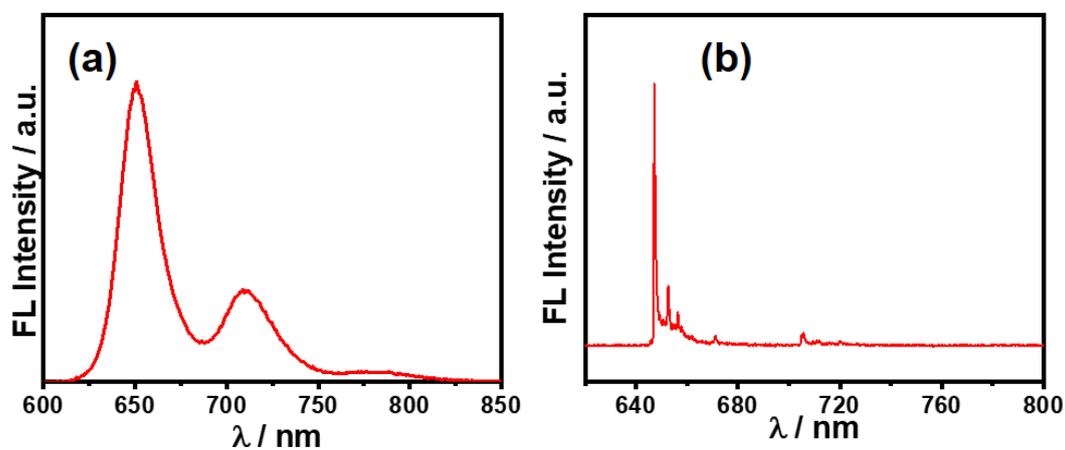

**Supplementary Fig. 10|** Fluorescence spectra of single TDI monomers at room temperature (a) and at T = 1.4 K (b).